\documentclass[sn-mathphys]{sn-jnl}% Math and Physical Sciences Reference Style
\jyear{2021}%

\theoremstyle{thmstyleone}%
\theoremstyle{thmstyletwo}%

\theoremstyle{thmstylethree}%

\begin{document}

\title[Article Title]{A scalable pipeline for COVID-19: the case study of Germany, Czechia and Poland.}


\author*[1,2]{\fnm{Wildan} \sur{Abdussalam}}\email{w.abdussalam@hzdr.de}

\author[1,2]{\fnm{Adam} \sur{Mertel}}

\author[1,2]{\fnm{Kai} \sur{Fan}}

\author[1,2,4]{\fnm{Lennart} \sur{Sch\"uler}}

\author[1,2]{\fnm{Weronika} \sur{Schlechte-We\l{}nicz}}

\author*[1,2,3,5]{\fnm{Justin M.} \sur{Calabrese}}\email{j.calabrese@hzdr.de}

\affil[1]{\orgdiv{Center for Advanced Systems Understanding}, \orgname{Helmholtz-Zentrum Dresden-Rossendorf e.V. (HZDR)}, \orgaddress{\street{Untermarkt 20}, \city{Görlitz}, \postcode{02826}, \state{Sachsen}, \country{Germany}}}

\affil[2]{\orgdiv{Helmholtz-Zentrum Dresden-Rossendorf}, \orgaddress{\street{Bautzner Landstrasse 400}, \city{Dresden}, \postcode{01314}, \state{Sachsen}, \country{Germany}}}

\affil[3]{\orgdiv{Department of Ecological Modelling},
\orgname{Helmholtz Centre for Environmental Research (UFZ)},\orgaddress{ \city{Leipzig}, \state{Sachsen}, \country{Germany}}}

\affil[4]{\orgdiv{Department of Computational Hydrosystems},
\orgname{Helmholtz Centre for Environmental Research (UFZ)},\orgaddress{ \city{Leipzig}, \state{Sachsen}, \country{Germany}}}

\affil[5]{\orgdiv{Department of Biology},
\orgname{University of Maryland},\orgaddress{\street{College Park}  \city{MD}, \country{USA}}}

\abstract{Throughout the coronavirus disease 2019 (COVID-19) pandemic, decision makers have relied on forecasting models to determine and implement non-pharmaceutical interventions (NPI). In building the forecasting models, continuously updated datasets from various stakeholders including developers, analysts, and testers are required to provide precise predictions. Here we report the design of a scalable pipeline which serves as a data synchronization to support inter-country top-down spatiotemporal observations and forecasting models of COVID-19, named the \textit{where2test}, for Germany, Czechia and Poland. We have built an operational data store (ODS) using PostgreSQL to continuously consolidate datasets from multiple data sources, perform collaborative work, facilitate high performance data analysis, and trace changes. The ODS has been built not only to store the COVID-19 data from Germany, Czechia, and Poland but also other areas. Employing the dimensional fact model, a schema of metadata is capable of synchronizing the various structures of data from those regions, and is scalable to the entire world. Next, the ODS is populated using batch Extract, Transfer, and Load (ETL) jobs. The SQL queries are subsequently created to reduce the need for pre-processing data for users. The data can then support not only forecasting using a version-controlled Arima-Holt model and other analyses to support decision making, but also risk calculator and optimisation apps~\cite{davoodi_modeling_2022,davoodi_optimal_2022}. The data synchronization runs at a daily interval, which is displayed at https://www.where2test.de.}




\maketitle

\section{Introduction}\label{sec:intro}
In building forecasting models of COVID-19, many researchers employ the training datasets provided by each country's representative institutions, e.g., Robert Koch Institute in Germany. The publicly accessible COVID-19 data provided in raw textual format, such as CSV, JSON, and XML are downloaded and analysed by the researchers employing either statistical or machine learning approaches. However, the data are unwell structured and require heavy pre-processing as well as ingestion activities for further analysis. This method is inherently inefficient due to identical and manual parallel pre-processing of the RKI data (using e.g. python or R scripts) performed by each researcher. This reduces the efficiency of each and everyone's work as all have to spend hours and days in pre-processing data before coming to modeling and forecasting. Advanced computing infrastructures and novel software pipelines are crucial tools to synchronize the data structures which originate from various sources and to extremely reduce heavy pre-processing~\cite{raisaro_scor_2020}. They serve as essential prerequisites to realise the data surveillance and outbreak response management, which have been implemented in fighting other endemic diseases~\cite{wangenheim_integrating_2019,fahnrich_surveillance_2015,smith_intermine_2012,pfander_scalable_2011}. 

To date, the data management has been applied in controlling the outbreak of COVID-19~\cite{kostkova_data_2021,budd_digital_2020,binti_hamzah_coronatracker_2020,centre_european_2022,naqvi_covid-19_2021,eudata_covid19-eu-data_2020,latinoamerica_latin_2020,agapito_covid-warehouse_2020,arora_serotracker_2021,govuk_interactive_2022,maryland_coronavirus_2022,rki_robert_2022,dresden_corona-dashboard_2022,dong_interactive_2020,sha_spatiotemporal_2021}. Most of them provide maps and the prevalent data in the following regional level: (i) National level, e.g., COVID-19 data  of World wide~\cite{binti_hamzah_coronatracker_2020}, Europe~\cite{centre_european_2022,naqvi_covid-19_2021,eudata_covid19-eu-data_2020}, and Latin America~\cite{latinoamerica_latin_2020}; (ii) State and county levels, e.g., the COVID-19 data warehouse for Italy~\cite{agapito_covid-warehouse_2020}, COVID-19 dashboard for UK~\cite{govuk_interactive_2022}, the COVID-19 dashboard for Maryland~\cite{maryland_coronavirus_2022}, and for Germany~\cite{rki_robert_2022}.; (iii) County level, e.g., Dresden, Germany~\cite{dresden_corona-dashboard_2022}. More completed version is provided by the John Hopkins University~\cite{dong_interactive_2020}, which serves the dashboard and the prevalent data for each regional level in the USA as well as for most of countries around the world. Likewise, the similar method in the presence of semi-automatic validation strategy was conducted to check the data quality of daily updated numbers with governmental/official data sources~\cite{sha_spatiotemporal_2021}. However, most of dashboards and data warehouses have not provided the features to let the users perform an inter-country top-down spatiotemporal observation, i.e., observing the inter-country prevalence and simultaneously being able to observe to the microscopic level (nation $\rightarrow$ state $\rightarrow$ county $\rightarrow$ municipality). The features could provide insights, for example, to study COVID-19 border dynamics which have been so far attracted considerable attentions~\cite{han_xiaoyi_quantifying_2021,laroze_covid-19_2021,grimee_modelling_2021,hossain_effects_2020}.  Moreover, they are lack of forecasting features, which play a key role in predicting the future prevalence as well as determining non pharmaceutical interventions (NPI). A tremendous number of forecasting models, e.g., agent-base \cite{liu_model-based_2022}, machine learning~\cite{bastani_efficient_2021,flaxman_estimating_2020}, combination model~\cite{haug_ranking_2020,liu_visitor_2021}, compartment model~\cite{lai_effect_2020,fanelli_analysis_2020,bertozzi_andrea_l_challenges_2020,schuler_data_2021,rahimi_review_2021}, time series~\cite{salgotra_time_2020,roy_spatial_2021,geng_changes_2021,wang_estimating_2021,sharma_modeling_2020,sahai_arima_2020,benvenuto_application_2020} have employed government datasets to provide essential inputs for public decisions. However, most of datasets that were used in those studies are limited to the specific time window which are likely to produce different results when the datasets are updated. Establishing a system of daily-updated-datasets assisted forecasts, therefore, is an alternative to improve their consistency and precision.

In this paper, we address the aforementioned issues by proposing the design of a scalable pipeline which allow us to perform the top-down spatiotemporal observation among Germany, Czechia, and Poland as well as to perform daily forecasts. The method of the pipeline which consists of extraction of various data sources and the ODS is described in subsec~\ref{subsec:pipeline}. More specifically, we will describe the dimensional fact database model and a daily migration process which underline the data synchronization between various data sources and our database server. We employ the dimensional fact model due to more flexibility and versatility in building spatiotemporal aggregation functions than the nanocubes model~\cite{l_lins_nanocubes_2013,bosworth_data_1995}. Next, in subsec~\ref{subsec:forecast} we will describe the time-series forecasting models which are supported by the presence of the ODS. Furthermore, the automatic system of daily forecasts owing to the presence of the pipeline will be laid out in this sub section. In Sec.~\ref{sec:results}, we will describe facilities that have been established due to the presence of the ODS. In order to demonstrate the inter-country top-down spatiotemporal observations, an analysis will begin from the macroscopic scale in which the study of the virus spread across the national borders is described in subsec~\ref{subsec:GIS}. Herein we consider the border among Germany, Czechia and Poland as a study case. In subsec~\ref{subsec:forecastresult}, we explore more microscopic level by applying a daily-updated-datasets assisted forecast for the prevalence in the state of Saxony, Germany. Last but not least, in subsec~\ref{subsec:superspreading}, most microscopic level that we will demonstrate is a superspreading event at a slaughter house in G\"utersloh, Lower Saxony, Germany. As the COVID-19 situation begins to enter an endemic phase, a study of superspreading event will provide essential information to trace the COVID-19 transmission after a mass event.
	
\section{Methods}\label{sec:method}
\subsection{Data Pipeline}\label{subsec:pipeline}
\begin{figure}[h]
\centering
\includegraphics[width=0.9\textwidth]{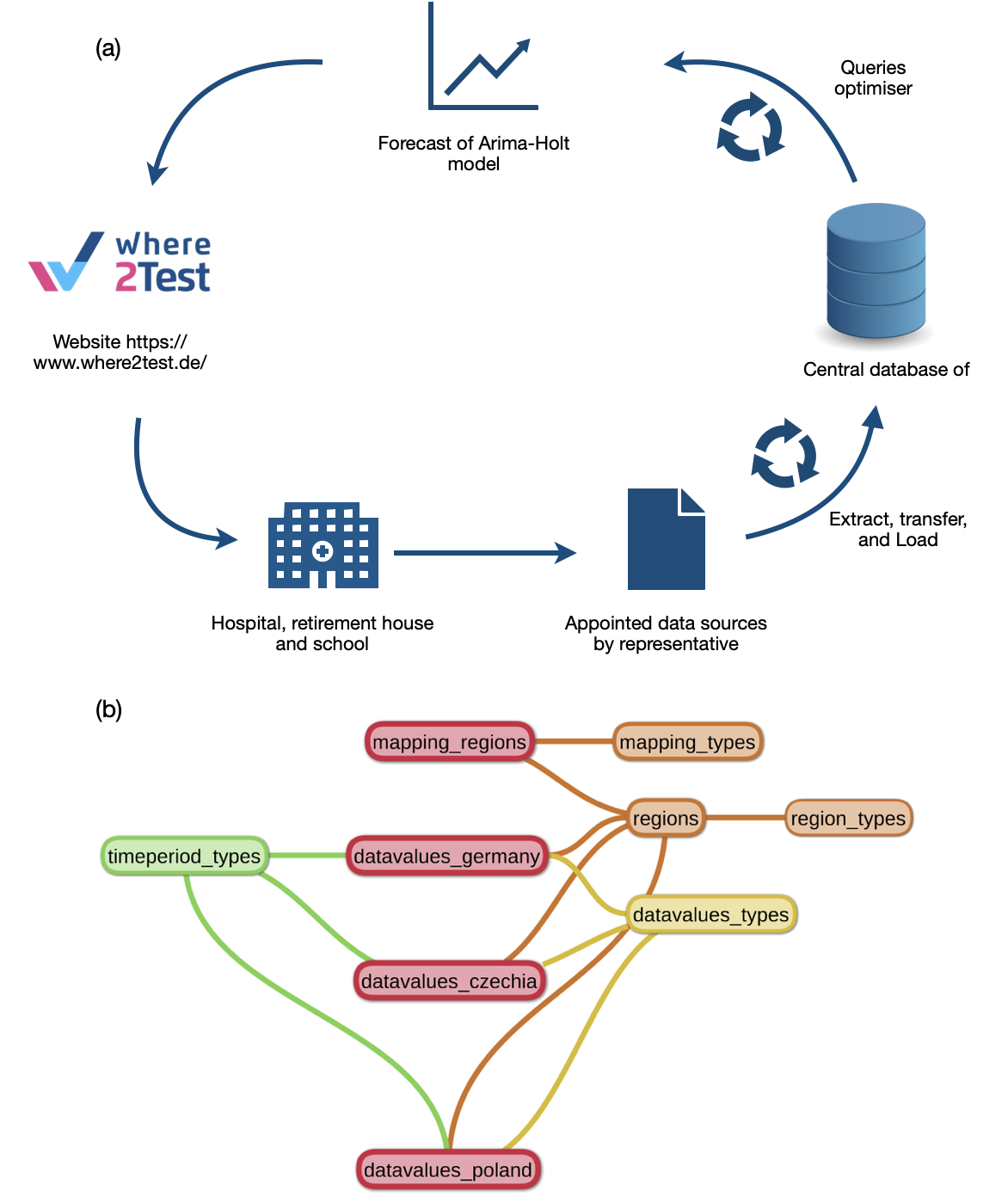}
\caption{\textbf{(a) A workflow of data pipeline} Hospitals, retirement houses, and schools of Germany, Czechia and Poland update the data of COVID-19 cases, vaccines and tests to the representative government institutions. A daily automatic ETL step is performed to synchronize the data sources and central database of CASUS. A daily and weekly automatic forecast employing, e.g. Arima-Holt model, is applied to provide rapid predictions. The predictions and the actual data are shown in the where2test website; (b) The scalable dimensional fact model. Datavalues and datavalue types represent \textbf{measures}, while region types and timeperiod types represent spatial and temporal \textbf{dimensions}, respectively.}\label{fig:pipeline}
\end{figure}
 
Fig.~\ref{fig:pipeline}a shows a workflow of the data pipeline. The hospitals, retirement houses and schools register the daily number of the COVID-19 cases and vaccines to the representative government. In order to consolidate these data, the relational database is built based on dimensional fact model~\cite{golfarelli_dimensional_1998}. Having established the relational database, the daily automatic extract, transfer and load (ETL) step is performed to migrate and integrate the data sources to the PostgreSQL database of CASUS HZDR (see Suplementary materials~\ref{subsec:dataworkflow}). Next, we create SQL inquiries-based views to be analysed by our researchers using the forecast and machine learning methods. The tested and completed analysis methods are set in the master stage and the other tested methods are set in the develop stage. Only the forecasting method in the master stage is integrated in the automatic pipeline.

The dimensional fact model is shown in Fig.~\ref{fig:pipeline}b. The model consists of three main concepts: (i) \textbf{Facts}, that refer to a subject of study (e.g., the study of infected, dead, recovered, hospitalised, test and vaccinated cases due to COVID-19); (ii) \textbf{Measures}, that refer to the quantitative data of the concept (\textit{i}). The measured data are stored in the table of datavalues. The tables of datavalues contain the number of infected, dead, recovered, hospitalised, test, and vaccinated cases due to COVID-19 in a given time and place. To date, the schema consists of three datavalues, i.e., datavalues of Germany, Czechia and Poland; (iii) \textbf{Dimensions}, that refer to temporal and spatial attributes. As the measured data are provided in a given time and place, the table of time period types and regions is necessary. The former stores the type of time period which consists of day and week data type; and the latter stores the necessary information of regions which consist of the name, abbreviation, ID of regions, ID of region type, geometry and population. The table of regions depends on the table of region types. The regions are categorised based on their sizes. The order of ascending sizes starts from municipality, county, state and nation. For Germany, the order of region type starts from Gemeinde, Kreise and Bundesland. Similar to Germany, Poland consist of Gmina, Powiat, and Wojewodztwo. Different from Germany and Poland, Czechia consist of 4 level, Obec, Orp, Okres and Kraj. The spatial and temporal attributes are connected by means of hierarchies to represent a -to-one relationship between them. The table of mapping$\_$types contains the hierarchical type of the spatial attributes, e.g., for Germany (Gemeinde to Kreise, Kreise to Bundesland), for Czechia (Obec to Orp, Orp to Okres and Okres to Kraj), and Poland (Gmina to Powiat and Powiat to Wojewodztwo). Next, a many-to-one relationship between those spatial hierarchies are stored in the table of mapping$\_$regions. Moreover, the table of timeperiod$\_$types consists of the hierarchical type of the temporal attributes.      

Aggregation functions are applicable on the measures along the temporal and spatial dimensions. For the former dimension, the weekly data are cumulative 7--day data. For example, a 7--day case reported on 13.03.2022 is an accumulation of the daily case for 07-13.03.2022. Moreover, for the latter dimension, county data are cumulative-municipality data. Not only accumulating the data from the municipality to a county level, in the presence of mapping regions table, it is possible to accumulate the data from the county to the state level as well as the state to the nation level. This allows us to scale the pipeline to other areas provided that the data of municipality are available from the sources.

\subsection{Forecasts}\label{subsec:forecast}
We employ auto regression integrated moving average (ARIMA) and Holt's linear trend models to forecast the infected, test, and hospitalised data of COVID-19 for Saxony (Germany), Czechia, and Poland. The ARIMA model has been successfully employed in predicting other endemic diseases~\cite{nsoesie_forecasting_2021,chen_avian_2019,he_epidemiology_2018,zeng_time_2016}. The model features suitable prediction based on time analysis series which is capable of providing short horizon forecast for most COVID-19 cases around the world~\cite{roy_spatial_2021,geng_changes_2021,wang_estimating_2021,sharma_modeling_2020,sahai_arima_2020,benvenuto_application_2020}. To make the model consistent and avoid overfitting, the order parameter of the ARIMA model is fixed instead of using the auto ARIMA model. The ARIMA is improved by employing the Holt’s linear trend model~\cite{holt_forecasting_2004}. The Holt’s model uses the exponential smoothing method to compute the weighted average of the past observation data~\cite{hyndman_forecasting_2018}. The forecasts from the Holt’s linear model have a trend, so the damped parameter is turned on to avoid this trend~\cite{gardner_why_2011,gardner_forecasting_1985,hyndman_forecasting_2018}. A self-defined mix function is used to compute the probability parameter m to combine the forecasts from two models and minimize the error. The Box-Cox transformation is used to normalize the input data~\cite{guerrero_time-series_1993,hyndman_forecasting_2018}. 

Our model provides a weekly forecast at first. In order to improve the daily variation and provide more real-time forecasts, we have built a daily forecast model. As the daily data have a clear weekly variation, the seasonal parameters are added to the model; and seasonal ARIMA (SARIMA) and Holt-Winters’ seasonal model are employed for the daily forecasts~\cite{hillmer_arima-model-based_1982,holt_forecasting_2004,winters_forecasting_1960}. Similar to the ARIMA model, the seasonal ARIMA model uses the fixed order and seasonal parameters. After comparing the errors from multiple methods, the additive method is selected for the Holt-Winters’ seasonal model. The mix function is also used for the daily forecasts to combine the forecasts from two models and improve the forecasting accuracy. For study cases of (S)Arima-Holt model, in Sec.~\ref{subsec:forecastresult}, we will provide the number of infections for Saxony, Germany. In addition to (S)ARIMA-Holt model, we employ outlier detection to identify and quantify Superspreading events. As suggested in~\cite{schuler_data_2021}, we identify and quantify superspreading events by using time series analysis based outlier detection methods. The rate of newly infected is modeled by an appropriate model, which could be something as simple as a rolling average to more elaborate ones as SIR-based models. The residues of the reported cases is used to identify outliers. At the same time, the residues can be used to quantify the size of a superspreading event.

\section{Results}\label{sec:results}
The presence of the pipeline has allowed us to provide following facilities: (i) The released data hub for dead and infected cases of all counties and states in Germany~\cite{abdussalam_post-processing_2022}, which allows a collaboration between CASUS research staffs and other external collaborators. The post-processing data serve as the clean data of daily infected and dead cases for county and state levels. In addition, we have also pre-processed the vaccination and hospitalization data for the county and municipal levels; (ii) The daily updated value of background risk for optimisation~\cite{davoodi_modeling_2022} and risk calculator apps~\cite{davoodi_optimal_2022}, which defines the chance of an average person who lives in the focal area, and carries out daily activities, will be infected over a one week period; (iii) Blog posts which update current COVID-19 situations in Germany. An interesting example of the posts would be the relation between the vaccination rate and the 7-day incidence in all states of Germany~\cite{mertel_where2test_2022}; (iv) Forecast- and model-based analysis. We explore the study cases mentioned in Sec.~\ref{sec:intro}, and begin by investigating of the virus spread across the national borders of Germany, Czechia, and Poland.

\subsection{Analysis of the virus spread across the national borders}\label{subsec:GIS}
COVID-19 spread among people. Therefore, human mobility is one of the most important factors defining the trend of spatiotemporal spreading of the virus. Understanding human mobility allows us to predict the spatiotemporal character of spread, evaluate the government steps restrictions, and provide effective non-pharmaceutical interventions. Primarily due to the heterogeneity of the sources and the interest scope of the particular research groups and communities, most of the COVID-19 research stays within the boundaries of one country. While most human mobility happens in the extent of one country or region, notably in Europe, the national border's mitigating effect is generally diminishing. To study the impact of the national border, several research papers \cite{eckardt2020covid,grimee2021modelling} applied various methodologies of geostatistics and geospatial modeling. More thorough quantification of the effect of border presence and international mobility on the epidemy requires a data storage integrating heterogeneous datasets across more countries.

The presented ODS infrastructure offers a possibility to study the spatiotemporal character of the virus spread on more levels, considering the effect of the national border. First, for our case study comprising the countries of Germany, Poland, and Czechia, we explored the correlation of new cases in the region, the distance and the border presence. We observed that the neighbour regions tend to have similar incidence values in the absence of barrier in the form of a national border among them. This step followed the research of McMahon et al. \cite{mcmahon2022spatial}, which showed a strong spatial autocorrelation of incidence values in the USA.

\begin{figure}[h]
\centering
\includegraphics[width=0.9\textwidth]{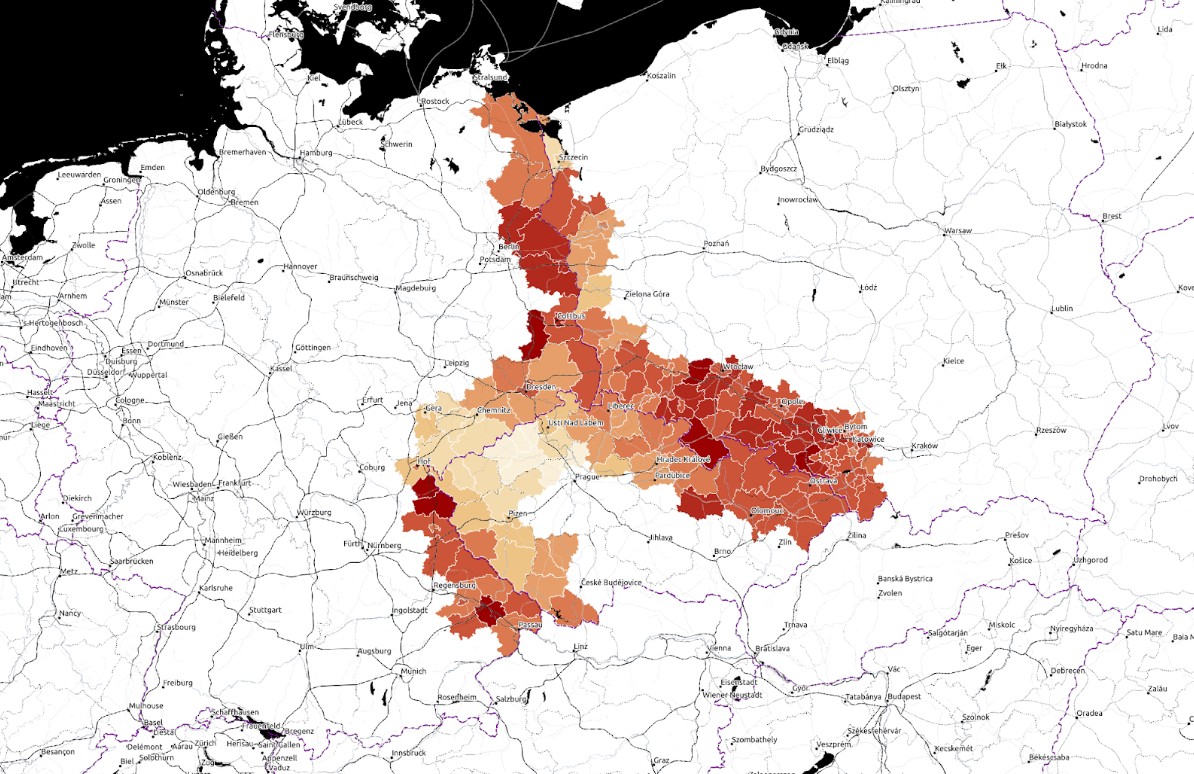}
\caption{Difference in the pair-wise correlations for regions within a 100 kilometer radius inside and outside the country. The red color represents the regions with the strongest difference, indicating the spread of the virus across the national borders.}\label{fig:border-correlations}
\end{figure}

Further, we calculated the average time-lagged pair-wise correlations for each region considering the regions in the radius of 100 kilometers, (i) within the same country, (ii) outside this country. The difference of these values can be seen in Fig.~\ref{fig:border-correlations}. The bigger difference represents regions where the incidence correlates much better than the regions within the same country, indicating a strong national border effect on the virus spread.

In the next step \cite{mertel2022fine}, we quantified the mitigation effect of the national border in more detail. We picked the state of Saxony in Germany and the neighboring regions in Czechia. For both countries, we collected and integrated the incidence data on the level of single municipalities. For each municipality, we constructed a local regression model which estimated the effect of three parameters, (i) border presence, (ii) municipality size, and (iii) temporal distance from other municipalities, on the spread of the virus. Based on this model, we identified very small-scale areas susceptible to a more intensive inter-national spread of the COVID-19.  

The top-down approach we selected for the study on the national border effect is possible thanks to the scalability of the implemented dimensional-fact model. This principle allows the ODS to comprise various administrative levels and combine various relevant topics within the perspective of spacetime. 

\subsection{Weekly and daily forecast of Arima-Holt and Sarima-Holt}\label{subsec:forecastresult}
\begin{figure}[h]
\centering
\includegraphics[width=0.9\textwidth]{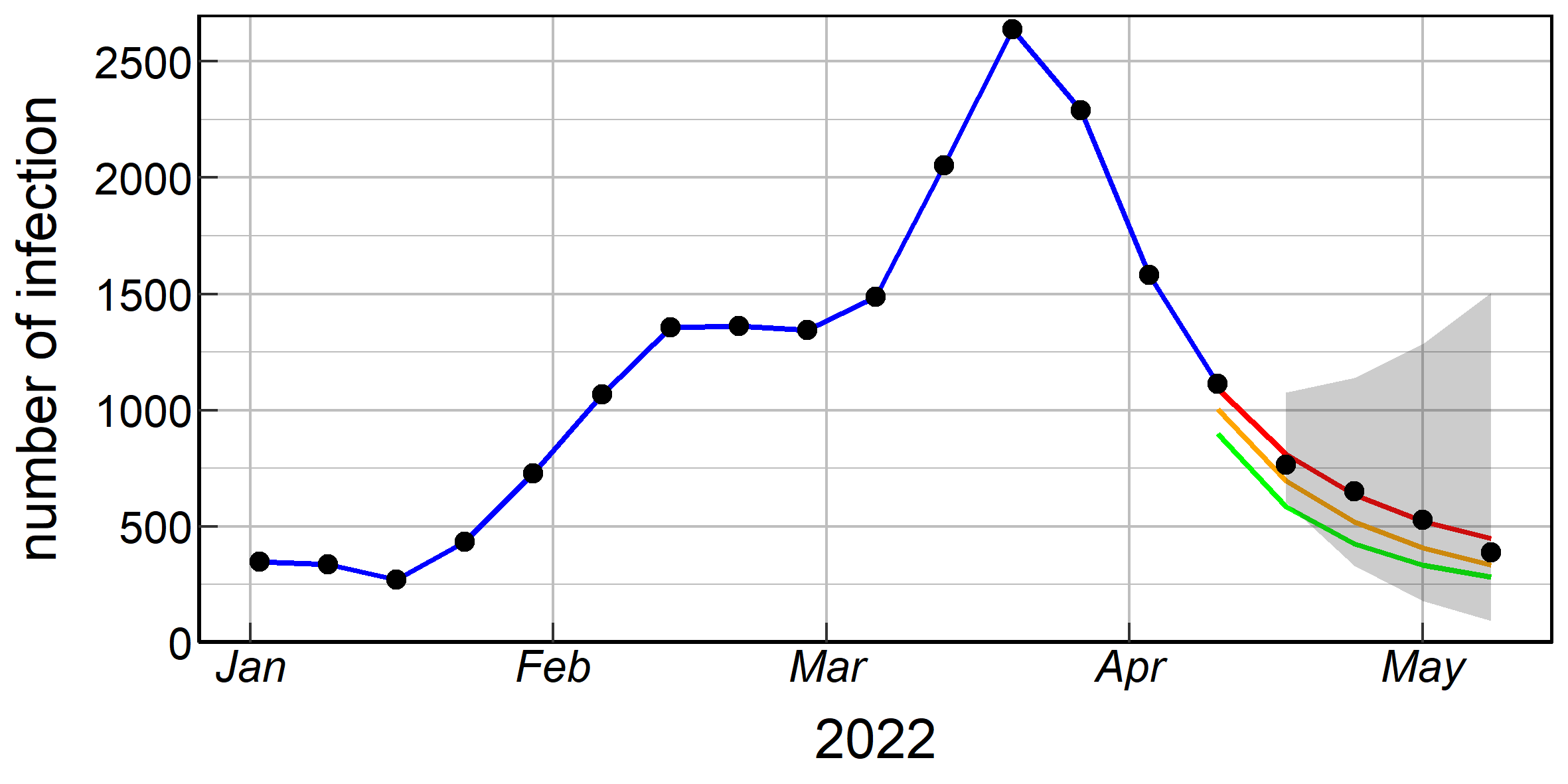} 
\caption{7-day incidence of infected cases Jan - 8 May 2022 for Saxony, Germany. The black dots denote the historical data, the blue line (\textcolor{blue}{---}) denotes a line guidance for the historical data, and the green (\textcolor{green}{---}), orange (\textcolor{orange}{---}), and red line (\textcolor{red}{---}) denotes the result of forecast using the Arima-Holt model performed on 10-04-2022, 11-04-2022, and 13-04-2022, respectively. The grey area shows the lower and upper limits of the forecast for 13-04-2022.}\label{fig:sachsenInfectedWeekly}
\end{figure}
For the case study, we provide a short-time forecast of 7-day incidence up to 4 horizons performed on 13-04-2022 using Arima-Holt model for Saxony, Germany. We used a training dataset of 13-04-2022 version which consists of the historical weekly data of Saxony and its counties from 01-03-2020 to 10-04-2022. The weekly data are automated-daily-updated data which are aggregated on Sunday (see Sec.~\ref{subsec:pipeline}). Although we update the data daily, for the case of Germany, the current and previous-day data are unavailable. In addition, the previous third day data are still to be updated from the source. When the forecast was performed on Sunday 10-04-2022, the number of infection on that day was less than the number of the same day for the following-day version. As a result, this produces inaccurate forecast (see Fig.~\ref{fig:sachsenInfectedWeekly}). As the day elapsed, more cases were automatically added and aggregated to the last Sunday data. Consequently, the performed forecast on 13-04-2022 provides higher exponent than the one with the dataset version of 10 and 11-04-2022. Moreover, the dataset of Wednesday consists of relatively-stable version. Therefore, the forecast is performed every Wednesday due to the consistency of data source for the last Sunday. 

\begin{figure}[h]
\centering
\includegraphics[width=0.9\textwidth]{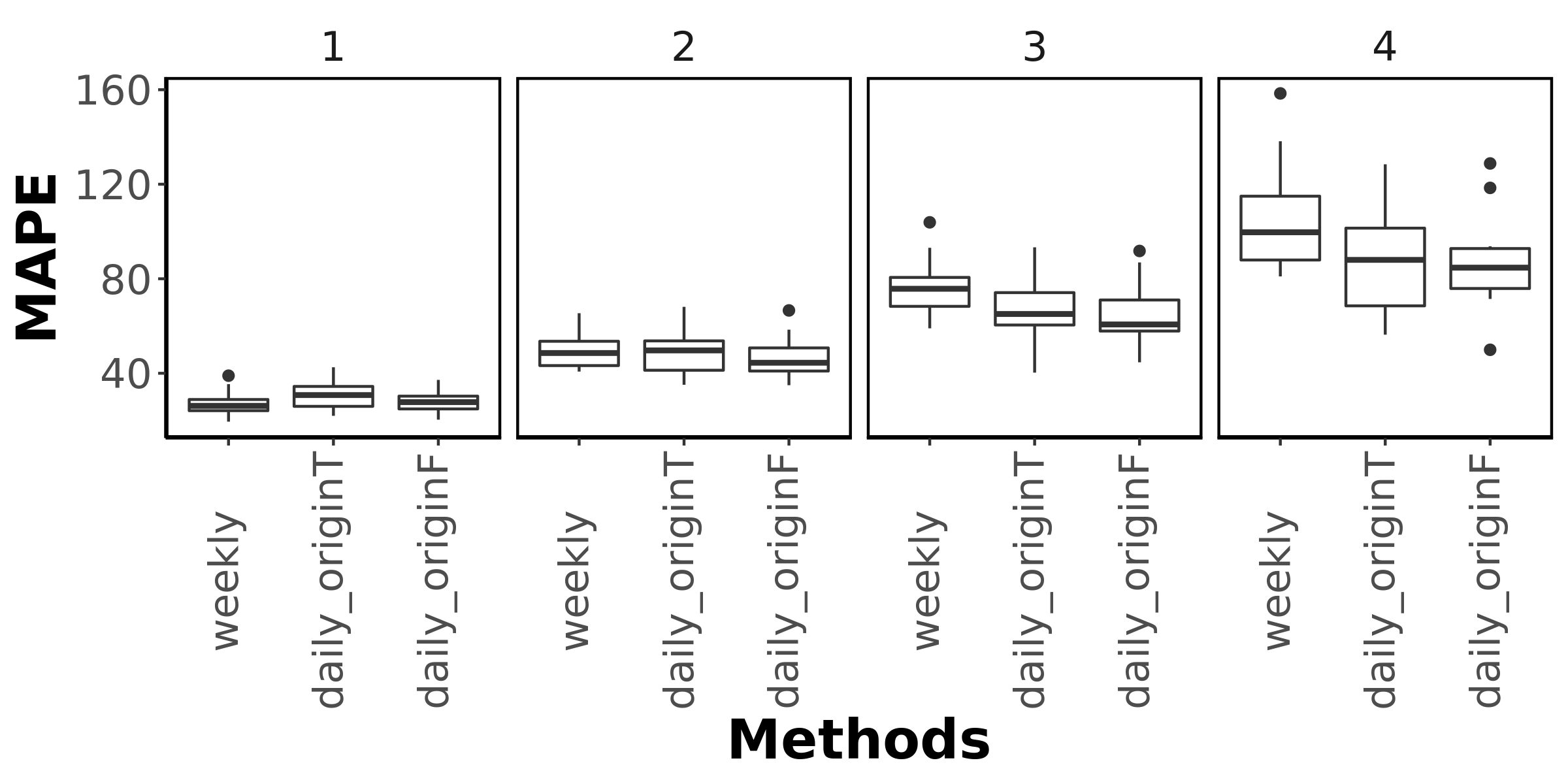} 
\caption{Mean absolute percentage error of Arima-Holt (weekly), Sarima-Holt in the presence of Box-cox transformation (daily\_originT), and Sarima-Holt in the absence of Box-cox transformation (daily\_originF) for 1$^{th}$ - 4$^{th}$ horizon.}\label{fig:arimaSarimaComparison}
\end{figure}

In order to check the four-horizon forecast, we compare it to the weekly-historical data updated on 11-05-2022. The latter consists of relatively stable data from 17-04-2022 to 08-05-2022. As shown in Fig.~\ref{fig:sachsenInfectedWeekly}, the weekly-historical data is surprisingly in quantitative agreement with the four-horizon forecast. However, this agreement occurs occasionally. When the forecast is performed in a different day, a deviation from the actual data for the following 4 horizons is likely to occur. Additional realisations of Arima-Holt forecast in Saxony and its counties, therefore, were performed to improve statistics. The realisations were performed every Wednesday from 05-01-2022 to 18-05-2022 in which the version-control dataset were employed as training and test datasets. An example would be a realisation of the Forecast on 05-01-2022. We used the weekly data version of 05-01-2022 as its training dataset and the weekly data version of the following 1st, 2nd, 3rd and 4th week as its test datasets. For each region, we then recorded a deviation of the forecast result from the historical data and quantified it as mean absolute percentage error (MAPE). As shown in Fig.~\ref{fig:arimaSarimaComparison}, the weekly Arima-Holt provides relatively low MAPE for the first and second horizon. For the third and fourth horizon, however, the range of MAPE tends to be wider than the first and second.

Therefore, we performed the Sarima-Holt model to improve the performance of forecast for the third and fourth horizon. Owing to daily-updated data, the version-control of daily data is employed as the seasonal parameters. In addition to the daily data, the Sarima-Holt forecast was performed using the same version-control weekly data employed to the Arima-Holt model. For the daily data, we removed the current and two previous-day data due to zero values for current and yesterday data, and inconsistent data for the previous third day. We then compared its performance in the presence and the absence of the Box-Cox transformation (BCT) used to normalize the input data. As shown in Fig.~\ref{fig:arimaSarimaComparison}, the Sarima-Holt model in the absence of the BCT provides lower MAPE than either the Arima-Holt or the Sarima-Holt in the presence of the BCT for not only the first and second horizons, but also the third and four horizons.  

\subsection{Superspreading events}\label{subsec:superspreading}
\begin{figure}[h]
\centering
\includegraphics[width=0.9\textwidth]{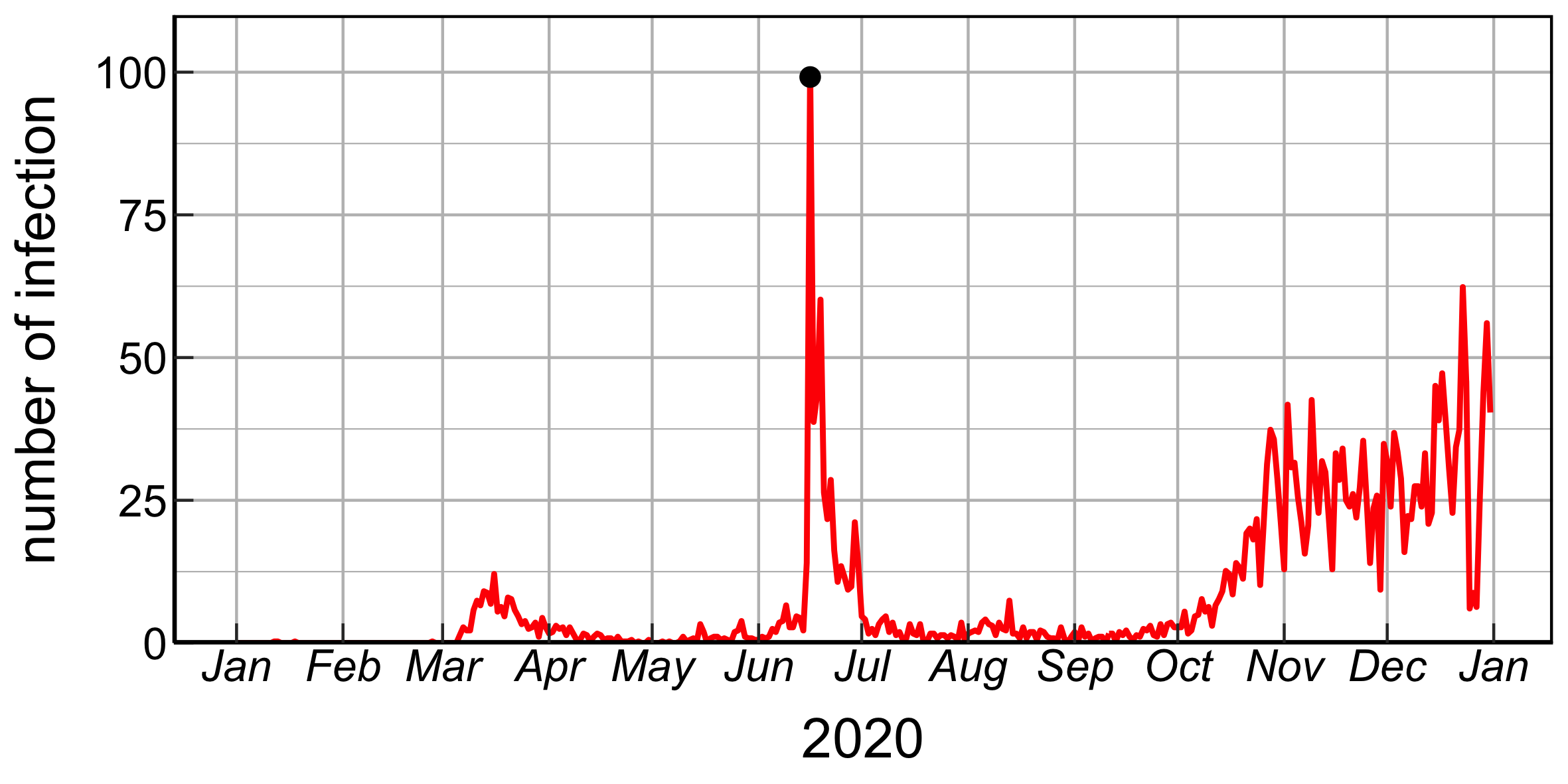} \\
\caption{The official reported COVID-19 daily incidence per 100.000 inhabitants in the district of Gütersloh. A superspreading event in a meat processing plant in June 2020 is successfully identified by an outlier detection method based on the Z-score (the black dot).}\label{fig:Outlier}
\end{figure}

Superspreding events play an important role in the dispersion dynamics of COVID-19~\cite{lemieux_phylogenetic_2020}. However, one of the most commonly used epidemiological model types, the compartment models, are not able to accuratly capture these events~\cite{schuler_data_2021, libotte_framework_2020}. We are currently working on a solution to the problem by using outlier detection methods on a county level. Many different methods exist and they can produce more robust results, when more than one timeseries is taken into account. A database as presented in this work is very advantageous, as it makes it very convenient to query the reported infections from all neighboring counties and use this additional data to more robustly identify outliers, which might be superspreading events. The largest confirmed superspreading to date in Germany with 1766 infections happened in a meat processing facility in the North Rhine-Westphalian district of G\"utersloh in June 2020. The facilities' environmental conditions combined with relatively close physical distance between workers were likely the main reason for efficient aerosol transmission~\cite{gunther_sars-cov-2_2020}. We take this event as an example to show the result of a Z-score based outlier detection method (Fig.~\ref{fig:Outlier}).

\section{Discussions}\label{sec:discussions}
Our analysis, implementing the pipeline in the presence of dimensional fact model has allowed us to daily migrate the data efficiently due to the functions of spatiotemporal aggregation. To provide the weekly data of counties, states, and nations, we only migrate the data of daily municipalities/counties (depends on the data availability of each nation) to the database server which are then aggregated to the higher spatiotemporal level. This model provides more advantages than the nanocubes model~\cite{l_lins_nanocubes_2013,bosworth_data_1995}. For the nanocubes model, each spatial (municipality, county, state and nation) and temporal (daily and weekly) data are required to be migrated to the database server. Consequently, this leads to a longer migration process than the one performed using the dimensional fact model. Moreover, its spatiotemporal mapping enables us to perform an efficient table join among national data which is confirmed by the application on the Subsec.~\ref{subsec:GIS}.  

The presence of daily-updated data due to the presence of the pipeline has allowed us to develop the Sarima-Holt model. The model shows more robust prediction for longer horizon than the Arima-Holt one. More specifically, the Sarima-Holt in the absence of the BCT outperforms the Arima-Holt model for the third and fourth horizon. This performance is due to the seasonal-parameter contribution to the model. As a result, the forecast tends to better predict for the third and fourth horizon. In contradiction, the Sarima-Holt in the presence of the BCT provides lower performance than the absence one due to less variation of the training data after BCT (see Fig.~\ref{fig:daily_cases_sachsen}). The Sarima-Holt model is trained by the daily data, and the variation of the data could make the model more sensitive to the infection change compared to the Arima-Holt model trained by the weekly data. However, the BCT reduces the variation of the daily data, and consequently the daily forecasts perform worse than in the absence of the BCT.

\section{Conclusion}\label{sec:conclusion}
Our work has demonstrated the utility of the data pipeline for top-down spatiotemporal analysis. We  have first shown the macroscopic analysis, in which the investigation of the virus spread across the national border is presented. At more microscopic level, we have demonstrated data-driven approach due to the presence of the pipeline which is applied to the prevalence of the county region. The daily-updated data has improved the precision of the model for longer horizon. This data-driven epidemic models provide more realistic forecast results than either the parsimonious~\cite{bertozzi_andrea_l_challenges_2020} or more number of parameters with agent-based method~\cite{liu_model-based_2022} due to the usage of daily-updated data. This may contribute to public health policy making, including contributing to public health forecasting teams. Last but not least, exploring to lower level of region, we have demonstrated that the outlier model is applicable to capture the superspreading event which occurred in 2020. These have shown that our work is capable of performing top-down analysis as well as rapid and precise forecasts due to the presence of the pipeline.    

\section{Data sources}\label{sec:data}
\begin{itemize}
    \item COVID-19 data for Germany, Czechia and Poland.
    \begin{itemize}
        \item Age-based hospitalisation of state level for Germany (\href{https://github.com/KITmetricslab/}{https://github.com/KITmetricslab/hospitalization-nowcast-hub/blob/main/data-truth/COVID-19/}).
        \item Age-based and types first, second, and third doses of vaccine for county level (\href{https://github.com/robert-koch-institut}{https://github.com/robert-koch-institut/COVID-19-Impfungen_in_Deutschland/blob/master/Aktuell_Deutschland_Landkreise_COVID-19-Impfungen.csv}).
        \item COVID-19 infected, recovered, hospitalised and dead cases of Dresden (\href{http://daten.dresden.de}{http://daten.dresden.de/duva2ckan/files/de-sn-dresden-corona_-_covid-19_-_fallzahlen_md1_dresden_2020ff/content}).
        \item COVID-19 infected, dead, and  test cases of Czechia for Municipality level (\href{https://onemocneni-aktualne.mzcr.cz}{https://onemocneni-aktualne.mzcr.cz/api/v2/covid-19/}).
        \item Age-based and gender-based infected and dead cases for county level of Germany (\href{Robert Koch Institute}{https://experience.arcgis.com/experience/478220a4c454480e823b17327b2bf1d4}).
        \item COVID-19 cases for municipality level of Saxony, Germany (\href{https://www.coronavirus.sachsen.de/}{https://www.coronavirus.sachsen.de/corona-statistics/rest/infectionOverview.jsp}).
        \item COVID-19 cases for county level of Saxony, Germany (\href{https://media.githubusercontent.com/media/robert-koch-institut}{https://media.githubusercontent.com/media/robert-koch-institut/SARS-CoV-2_Infektionen_in_Deutschland/master/Aktuell_Deutschland_SarsCov2_Infektionen.csv})
        \item COVID-19 infected, dead, and test cases for county level of Poland (\href{https://wojewodztwa-rcb-gis.hub.arcgis.com}{https://wojewodztwa-rcb-gis.hub.arcgis.com/pages/dane-do-pobrania}).
        \item COVID-19 vaccine for county level of Poland (\href{https://www.gov.pl}{https://www.gov.pl/web/szczepimysie/raport-szczepien-przeciwko-covid-19}).
        \item COVID-19 types in Sachsen (\href{https://www.coronavirus.sachsen.de}{ https://www.coronavirus.sachsen.de/infektionsfaelle-in-sachsen-4151.html}). 
    \end{itemize}
    \item Dictionaries of regions. 
    \begin{itemize}
        \item Administrative areas in Germany (\href{https://gdz.bkg.bund.de}{ https://gdz.bkg.bund.de/index.php/default/digitale-geodaten/verwaltungsgebiete.html}). 
        \item Administrative areas in Poland (\href{https://gis-support.pl}{https://gis-support.pl/baza-wiedzy-2/dane-do-pobrania/granice-administracyjne/})
        \item Administrative areas in Czechia (\href{https://geoportal.cuzk.cz}{https://geoportal.cuzk.cz/(S(1nhx02lray0vkrhce1y2d53d))/Default.aspx?mode=TextMeta&text=dSady_RUIAN&side=dSady_RUIAN})
        \item Population numbers in Czech municipalities (\href{www.czso.cz}{https://www.czso.cz/csu/czso/pocet-obyvatel-v-obcich-k-112021})
        \item Postal codes in Germany (\href{https://www.geonames.org}{https://www.geonames.org/postal-codes/postleitzahlen-deutschland.html})
        \item Population numbers in Poland ({\href{https://stat.gov.pl} {https://stat.gov.pl/obszary-tematyczne/ludnosc/ludnosc/ludnosc-stan-i-struktura-ludnosci-oraz-ruch-naturalny-w-przekroju-terytorialnym-stan-w-dniu-30-06-2021,6,30.html}})
    \end{itemize}
\end{itemize}

\section{Code availability}\label{sec:code}
The codes are publicly accessible on \href{Rodare}{https://www.hzdr.de/publications/Publ-34430}.

\backmatter

\section{Supplementary information}\label{sec:supinfo}
\subsection{Data workflow}\label{subsec:dataworkflow}
We use \href{Talend-Java}{https://www.talend.com/products/talend-open-studio/} to perform data migration. The migration between the data sources and the PostgreSQL database of CASUS HZDR has been performed as follows:

\begin{figure}[h]
\centering
\includegraphics[width=0.9\textwidth]{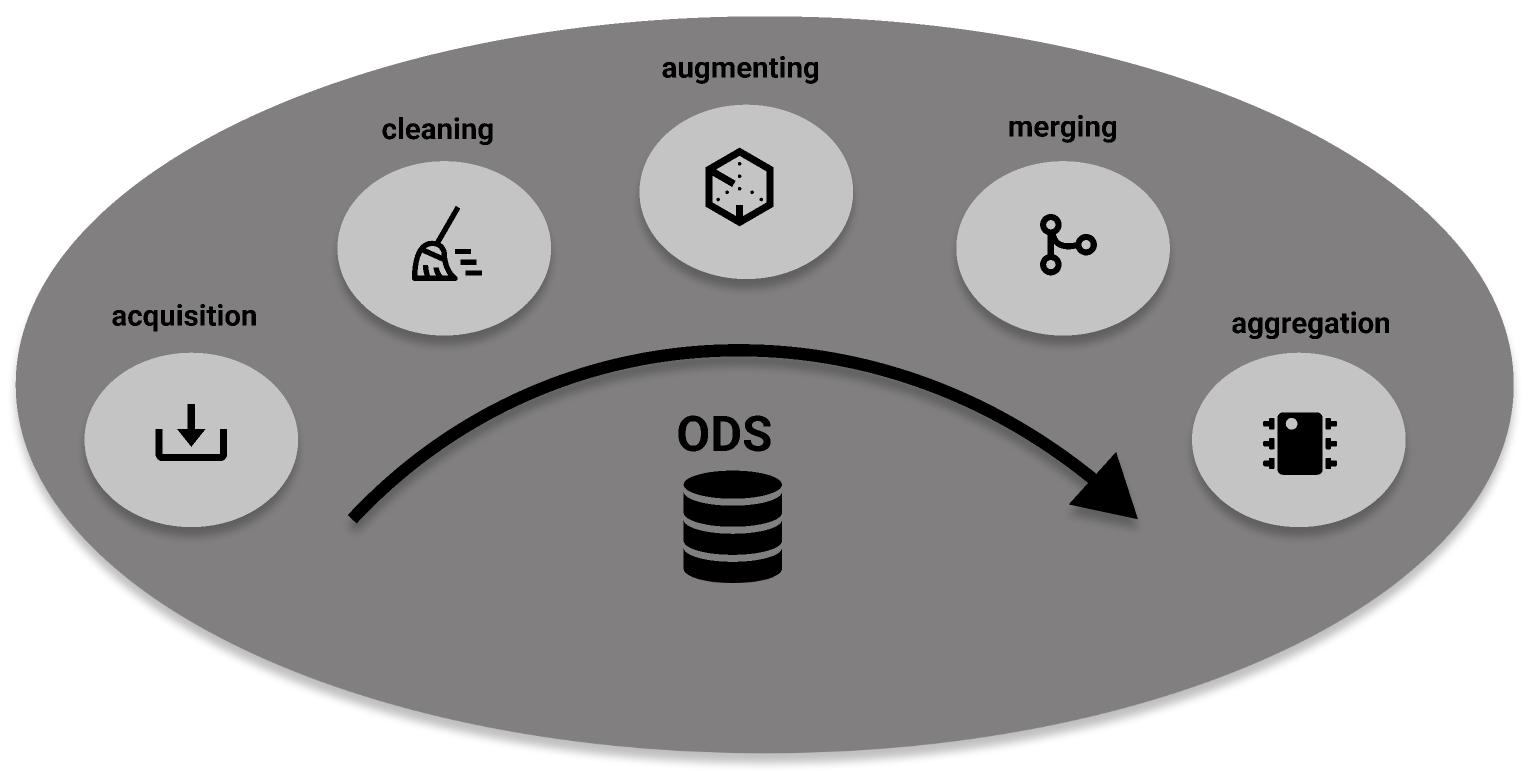}
\caption{\text{Data workflow of the ETL process (see texts for its description).}}\label{fig:dataWorkflow}
\end{figure}

\begin{enumerate}
    \item \textbf{Data acquisition}
    \newline
    The data are automatically downloaded from sources~\ref{sec:data}. They are subsequently stored on the repository of where2test server. The downloaded data serve as data inputs of a migration process.  

 \item \textbf{Dictionaries and data augmentation}
    \newline
	To integrate and further augment data from heterogeneous sources (various forms, schema, temporal and spatial extent), we needed to prepare a list of dictionaries. We formed a dictionary for each spatial level in every country to cover all regions in our datasets. Here we included the unique region id, all alternative names, full names, geometries, and population numbers. This concept can be further extended to other values such as socioeconomical parameters, and information about the region. This way we are able to maintain the consistency in all datasets and enable their integration process. The list of sources used for building the dictionaries can be found in section Data Sources~\ref{sec:data}.
    \item \textbf{Data cleaning}
    \newline
    We migrate first \textbf{timeperiod\_types}, \textbf{region\_types}, \textbf{datavalues\_types}, and \textbf{mapping\_types}. While migrating the data to those tables, primary key are automatically set by a transformator (The script which migrates the data to the postgreSQL database.). Next, the primary key of those tables serves as the foreign key of other tables following the table relation shown in Fig.~\ref{fig:pipeline}b. An example would be a table of \textbf{regions} which contains intrinsic ID set by representative governments. In order to differentiate ID among Germany, Czechia and Poland, we add 'DE', 'CZ', 'PL', respectively, followed by the intrinsic ID. For the table of \textbf{regions}, the primary key of \textbf{region\_types} serves as its foreign key. The intrinsic IDs are categorised based on the ID of region types. A specific example would be Dresden, whose the intrinsic ID 14162. After cleaning processes, the intrinsic ID will be DE 14162 and categorised to the state level of \textit{Kreise}.
    
    Having migrated the data to the aforementioned tables, the table of \textbf{mapping\_regions} is occupied by the spatial-relation data. It contains the foreign key of the mapping type ID. An example would be a county Dresden. Dresden are mapped onto the state of Saxony and categorized to the mapping type \textbf{Kreis\_To\_Bundesland}. Next, the table of datavalues for nations is occupied by the data input. The datavalues table consists of three foreign keys which originate from the tables of \textbf{timeperiod\_types}, \textbf{regions}, \textbf{datavalues\_types}. In the presence of these foreign keys, a data merging process is feasible, which is described on the following item.
   
	\item \textbf{Data merging}
    In addition to the aforementioned three-foreign keys, date is set as the fourth attribute which allow us to perform data merging through inner join of tables. The inner join is employed to cleanly merge and avoid duplicated data on the table of datavalues. For instance, daily infected data of the lowest-level region for period of date are migrated to the table of \textbf{datavalues\_germany}. When the data sources are updated, they sometimes update the cases of the elapsed date. Inner join method allows us to automatically update the value of the elapsed date by the latest value. Moreover, when the new data with the latest date are present from the source, it allows automatic addition of the data to the table.
   
 	\item \textbf{Data aggregation}
    The presence of daily data of the lowest regions allow us to perform both time and spatial aggregations. Using functions, the time aggregation from daily to weekly period is feasible. Moreover, as mentioned on the Sec.~\ref{sec:method}, the spatial aggregation from the low to the high region level is allowable in the presence of the \textbf{mapping\_regions} table. 

\subsection{Additional forecasting results}\label{subsec:aforecastresults}    
    \begin{figure}[h]
    \centering
    \includegraphics[width=0.9\textwidth]{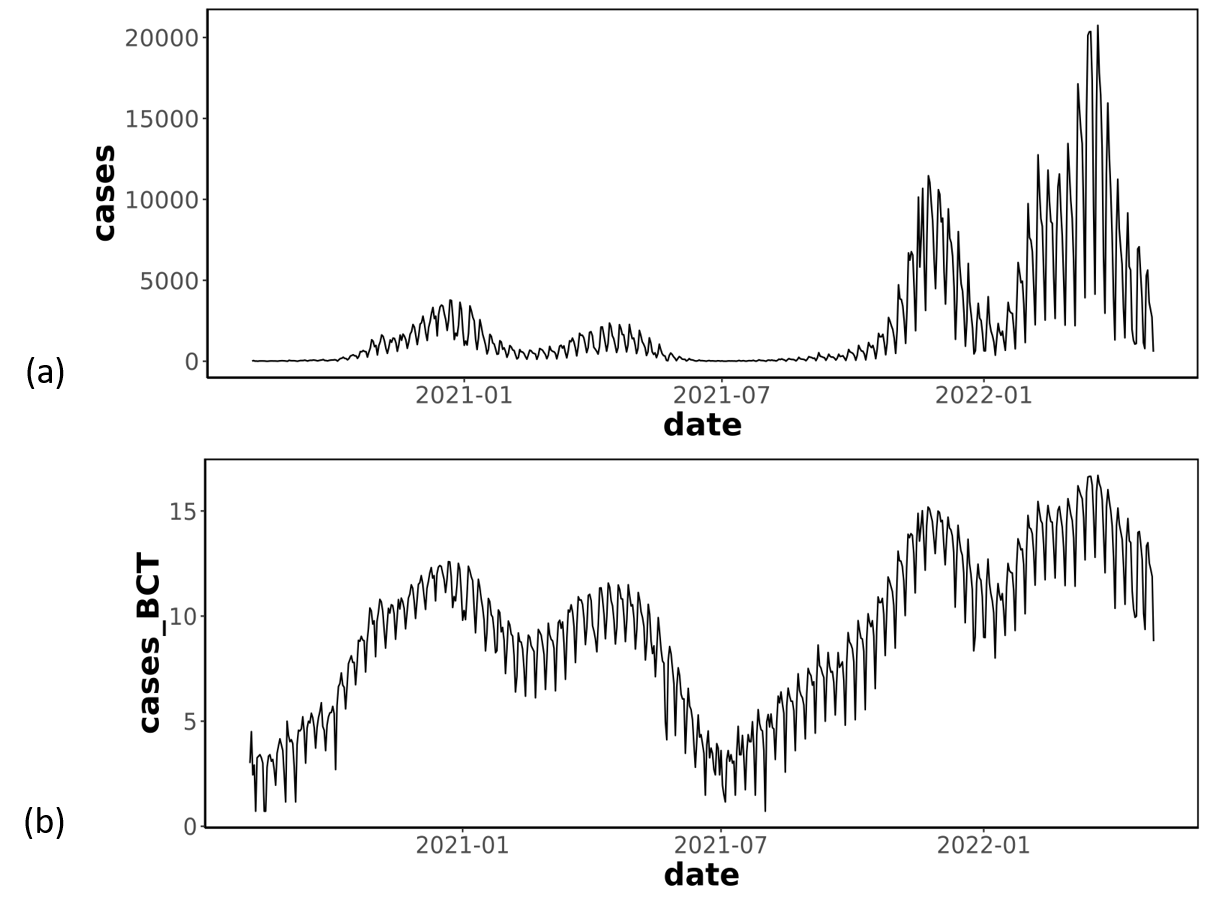}
    \caption{Time series of daily infected cases from Aug. 5, 2020 to Apr. 30, 2022 (a) before and (b) after Box-Cox transformation, respectively.}\label{fig:daily_cases_sachsen}
    \end{figure}
\end{enumerate}

\section*{Acknowledgments}
 This work was partially funded by the Center of Advanced Systems Understanding (CASUS), which is financed by Germany's Federal Ministry of Education and Research (BMBF) and by the Saxon Ministry for Science, Culture and Tourism (SMWK) with tax funds on the basis of the budget approved by the Saxon State Parliament. We thank to Jens Steiner for providing us virtual server of HZDR.
 
\section*{Author Contributions}
Study design: WA, AM, KF, LS, WSC and JMC; Study investigators:WA and JMC; Data acquisition: WSC, AM, and WA; Data Pipeline, ODS and automation: WA; Data analysis: WA, AM, KF, LS, JMC; Data interpretation: WA, AM, KF, LS, WSC and JMC; Manuscript preparation: WA, AM, KF, and LS; Manuscript review and revisions: WA, AM, KF, LS, WSC, and JMC; Final approval of manuscript: All authors.

\section*{Competing Interest}

The authors declare no competing interests.


\bibliography{epidemiology}% common bib file


\end{document}